\documentclass[epj]{svjour}
\usepackage{epsfig}
\usepackage{amssymb}
\newcommand{\lag}{\mathcal L}
\begin{document}
\title{Kaon condensation in neutron star 
using relativistic mean field models}
\author{S. W. Hong \inst{1} \and 
C. H. Hyun \inst{1}
\and
C. Y. Ryu \inst{2}
}                     
\institute{Department of Physics and Institute of Basic Science, 
Sungkyunkwan Univ., Suwon 440-746, Republic of Korea \and 
Research Center for Nuclear Physics (RCNP), Ibaraki, Osaka 567-0047,
Japan}
\date{Received: date / Revised version: date}
\abstract{
We use the modified quark-meson coupling 
and the quantum hadrodynamics models to
study the properties of neutron star.
Coupling constants of both models are adjusted to reproduce
the same saturation properties.
The onset of kaon condensation in neutron star matter
is studied in detail
over a wide range of kaon optical potential values.
Once the kaon condensation takes place, the population of
kaons increases very rapidly, and kaons become the dominant component,
possibly making the neutron star matter a kaonic matter
if the kaon optical potential is large.
\PACS{
      {97.60.Jd}{Neutron stars}   \and
      {14.20.Jn}{Strange particles}
     } 
} 
%
\maketitle
\section{Introduction}
\label{sec:introduction}

Observations of the masses of binary pulsars \cite{tc-apj99},
which are candidates of neutron stars indicate that the
maximum mass of neutron stars are roughly around $1.5 M_\odot$.
While the mass of a neutron star with only neutrons and protons 
is estimated to be about $2 M_\odot$, 
exotic degrees of freedom \cite{kpe-prc95,s-pal,nkg-prd92} 
other than the nucleons seem to reduce the 
maximum mass of a neutron star close to the observed values,
which implies that the
exotic degrees of freedom such as creation of hyperons,
Bose-Einstein condensation, strange matter, and quark deconfinement
seem to be needed to reproduce the observed masses of neutron stars.

In this work, we consider the strangeness degrees of freedom by including
hyperon creation and kaon condensation in the neutron star matter.
The equation of state (EoS) of dense nuclear matter is sensitive to 
the in-medium interaction of the hyperons and kaons.
Recently, the magnitudes of the kaon-nucleus potential
in matter have attracted much attention.
Some calculations \cite{Schaffner,oset20,Cieply}
show that the real part of the $K^-$-nucleus optical potential $U_{K^-}$
is shallow ($U_{K^-} \approx$ $-$50 MeV), but some other calculations
suggest that $U_{K^-}$ can be as large as about $-120$ MeV
\cite{gal94,Kaiser} or even close to $-200$ MeV \cite{Batty}.
We consider the possibility of deep optical potential of kaons in nuclei 
and explore the consequences in the composition of neutron star matter
and the mass-radius relation of the neutron star.

To treat the dense nuclear matter we employ two different relativistic
mean field models; the modified quark-meson coupling (MQMC) model \cite{mqmc}
and the quantum hadrodynamics (QHD) model \cite{sw-qhd}.
In the MQMC model the nucleons and hyperons in the baryon octet 
are treated as MIT bags, whereas in QHD they are assumed to be point particles.
The parameters of the two models are calibrated to produce exactly
the same saturation properties.
By comparing the results from the two different models we can
investigate the model dependence at high densities. 
We find the onset densities of the kaon condensation and the compositions
of matter at high densities are non-negligibly model dependent.
The mass-radius relation of the neutron star can significantly 
differ depending on the interaction of the kaon in nuclear matter.

\section{Models}
\label{sec:model}

The model Lagrangian consists of 
several terms for the octet baryons, exchange mesons, leptons and kaons;
$\lag_{tot} = \lag_B + \lag_M + \lag_l + \lag_K$.
In the mean-field approximation, each term reads
\begin{eqnarray}
\lag_B  &=& \sum_B \bar \psi_B \big[i\gamma \cdot \partial - m_B^*
(\sigma, \sigma^*) \nonumber \\ & &
-\gamma^0 \left(g_{\omega B}\omega_0 + g_{\phi B}\phi_0 + \frac 12 g_{\rho B}
\tau_z \rho_{03} \right) \big] \psi_B, \label{eq:lagb} \\
\lag_M &=& -\frac 12 m_\sigma^2 \sigma^2 - \frac 12 m_{\sigma^*}^2 {\sigma^*}^2
+ \frac 12 m_\omega^2 \omega_0^2 + \frac 12 m_\phi^2 \phi_0^2 \nonumber \\
& & + \frac 12 m_\rho^2 \rho_{03}^2, \label{eq:lagm} \\
\lag_l &=& \sum_l \bar \psi_l ( i \gamma \cdot \partial - m_l)\psi_l ,
\label{eq:lagl}, \\
\lag_K &=& D_\mu^* K^* D^\mu K - {m_K^*}^2 K^* K,
\label{eq:lagk}
\end{eqnarray}
where $B$ denotes the sum over the octet baryons
and $l$ stands for the sum over the free electrons and muons.
The interactions between the non-strange light quarks ($u$ and $d$)
are mediated by $\sigma$, $\omega$ and $\rho$ mesons, and
$\sigma^*$ and $\phi$ mesons are introduced to take into
account the interactions between $s$ quarks.
One of the most important differences between MQMC and QHD lies in
the treatment of baryons; the former treats baryons as composite systems of
quarks, and the latter assumes baryons as point particles.
As a result, the analytic form of the effective mass of a baryon differs
greatly as follows.

In MQMC, the effective mass of a baryon in matter 
$m^{*}_B(\sigma, \sigma^*)$ can be written as
$
m^*_B = \sqrt{E^2_B - \sum_q  \left(\frac{x_q}{R} \right)^2},
$
where $E_B$ is the bag energy of a baryon, $R$ the bag radius
and $x_q$ the eigenvalue of the quarks in the bag.
The bag energy of a baryon is given by
$
E_B = \sum_q  \frac{\Omega_q}{R} - \frac{Z_B}{R}
+ \frac{4}{3} \pi\, R^3\, B_B,
$
where $B_B$ and $Z_B$ are the bag constant and
a phenomenological constant for the zero-point motion of a baryon $B$,
respectively, and $\Omega_q = \sqrt{x^2_q + (R m^*_q)^2}$,
where $m^*_q (= m_q - g^q_\sigma \sigma - g^q_{\sigma^*} \sigma^*)$
is the effective mass of a quark.
In the MQMC model, the bag constant $B_B$ is assumed to depend on density,
and we use the extended form of Ref.~\cite{s-pal}.

In QHD, the effective mass of a baryon is written as
$
m^*_B = m_B - g_{\sigma B} \sigma - g_{\sigma^* B} \sigma^*.
$
We require MQMC and QHD to satisfy identically the same saturation 
properties. To produce the compression modulus within 
a reasonable range one needs to include
self-interaction terms of the $\sigma$-field
represented by
$
U^{\rm QHD}_{\sigma} = \frac{1}{3} g_2\, \sigma^3 +
\frac{1}{4} g_3\, \sigma^4
$
in the Lagrangian for QHD.
The meson term in the QHD thus reads 
$
\lag_M^{\rm QHD} = \lag_M - U^{\rm QHD}_{\sigma}.
$

The kaon is treated as a point particle in both MQMC and QHD models
in this work, and its effective mass is given as
$
m_K^* = m_K - g_{\sigma K} \sigma - g_{\sigma^* K} \sigma^*.
$
Covariant derivatives in Eq.~(\ref{eq:lagk}) include interactions 
with vector mesons through
$
D_\mu = \partial_\mu + i g_{\omega K}\omega_\mu
-i g_{\phi K} \phi_\mu + i \frac 12 g_{\rho K} \vec \tau \cdot \vec \rho_\mu.
$
A plane wave solution for the kaon field equation gives the dispersion relation
of the anti-kaon
\begin{eqnarray}
\omega_K = m_K^* - g_{\omega K} \omega_0
+ g_{\phi K} \phi_0 - g_{\rho K} \frac 12 \rho_{03}.
\label{eq:kaon-dispersion}
\end{eqnarray}

Nuclear saturation properties are relatively well known,
and the values quoted in the literatures lie in the range
$0.15 \sim 0.17$ fm$^{-3}$ for the saturation density $\rho_0$,
$15 \sim 16.3$ MeV for the binding energy per nucleon $E_b$,
$30 \sim 35$ MeV for the symmetry energy $a_{\rm sym}$,
$200 \sim 300$ MeV for the compression modulus $K$,
and $(0.7 \sim 0.8) m_N$ for the effective mass of the nucleon.
In this work we assume the saturation properties 
$\rho_0 = 0.17$ fm$^{-3}$, $E_b = 16.0$ MeV, $a_{\rm sym} = 32.5$ MeV,
$K = 285$ MeV, and $m^*_N = 0.78 m_N$. These conditions determine
the coupling constants of non-strange mesons to $u$ and $d$ quarks
in the MQMC, and to nucleons in the QHD.
These coupling constants are summarized in Table~\ref{tab:coupling}.
\begin{table}[t]
\caption{Non-strange meson coupling constants for the MQMC (left four
columns) and for the QHD (right four columns) models. $g_{\rho N}$ 
in QHD is the same as $g^q_\rho$ in MQMC.}
\label{tab:coupling}
\begin{center}
\begin{tabular}{cccc|cccc}
\hline\noalign{\smallskip}
$g^q_\sigma$ & $g^q_\omega$ & $g'^B_\sigma$ & $g^q_\rho$ &
$g_{\sigma N}$ & $g_{\omega N}$ & $g_2 ({\rm fm^{-1}})$ & $g_3$ \\
\noalign{\smallskip}\hline\noalign{\smallskip}
1.0 & 2.71 & 2.27 & 7.88 & 8.06 & 8.19 & 12.1 & 48.4 \\
\noalign{\smallskip}\hline
\end{tabular}
\end{center}
\end{table}

For the coupling constants between the non-strange mesons 
($\sigma$, $\omega$, and $\rho$) and hyperons 
we use the quark counting rule assuming that the $s$-quark 
does not couple to $u$ and $d$ quarks.
This gives us
$g^s_{\sigma} = g^s_{\omega} = g^s_{\rho} = 0,$
and these relations determine the coupling constants of
non-strange mesons to hyperons in both MQMC and QHD.
For the coupling constants of strange meson to hyperons, we assume
SU(6) symmetry, which gives us 
$g_{\sigma^*}^s = \sqrt 2 g_\sigma^{u,d}$ and 
$g_\phi^s = \sqrt 2 g_\omega^{u,d}$.

We have five meson-kaon coupling constants, $g_{\sigma K}$,
$g_{\omega K}$, $g_{\rho K}$, $g_{\sigma^* K}$ and $g_{\phi K}$.
The quark counting rule is employed for $g_{\omega K}$ and $g_{\rho K}$.
$g_{\sigma^* K}$ can be fixed from $f_0(980)$ decay, 
and $g_{\phi K}$ from the SU(6) relation
$\sqrt{2} g_{\phi K} = g_{\pi \pi \rho}$.
$g_{\sigma^* K}$ and $g_{\phi K}$ thus fixed are 2.65 and 4.27,
respectively.
The remaining coupling constant, $g_{\sigma K}$,
can be related to the real part of the optical potential
of a kaon at the saturation density through
$U_{K^-} = -(g_{\sigma K}\sigma + g_{\omega K}\omega_0)$.
The resulting $g_{\sigma K}$ values for several $U_{K^-}$ are given in
Table~\ref{tab:gsigmak}.

Once the coupling constants are determined, one can 
obtain the EoS of neutron star matter by solving self-consistently
the equation of motion of exchange mesons, 
charge neutrality condition, baryon number conservation, 
and $\beta$-equilibrium conditions for baryons and kaons.
The mass-radius relation of the neutron star can be obtained
by inserting the EoS into the Tolman-Oppenheimer-Volkoff (TOV) equation.
\begin{table}[b]
\caption{$g_{\sigma K}$ for several $U_{K^-}$ values
in MQMC and QHD.}
\begin{center}
\label{tab:gsigmak}
\begin{tabular}{cccc}
\hline\noalign{\smallskip}
$U_{K^-}$ (MeV) & $-120$ & $-140$ & $-160$  \\
\noalign{\smallskip}\hline\noalign{\smallskip}
MQMC  & 2.75 & 3.50 & 4.25 \\
QHD   & 2.83 & 3.61 & 4.39 \\
\noalign{\smallskip}\hline
\end{tabular}
\end{center}
\end{table}

\section{Results}
\label{sec:result}
\begin{figure}
\begin{center}
\epsfig{file=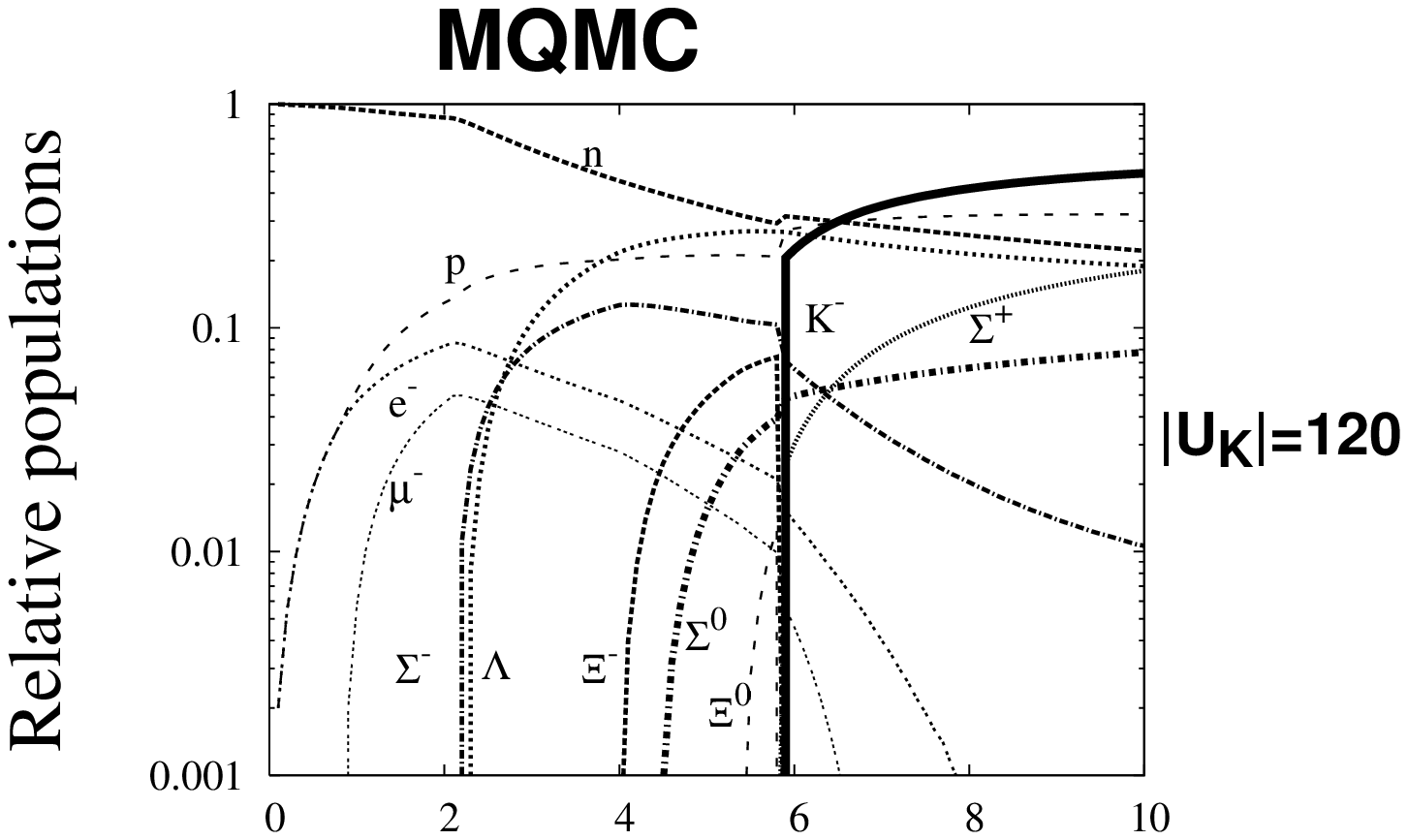, width=4.3cm}
\epsfig{file=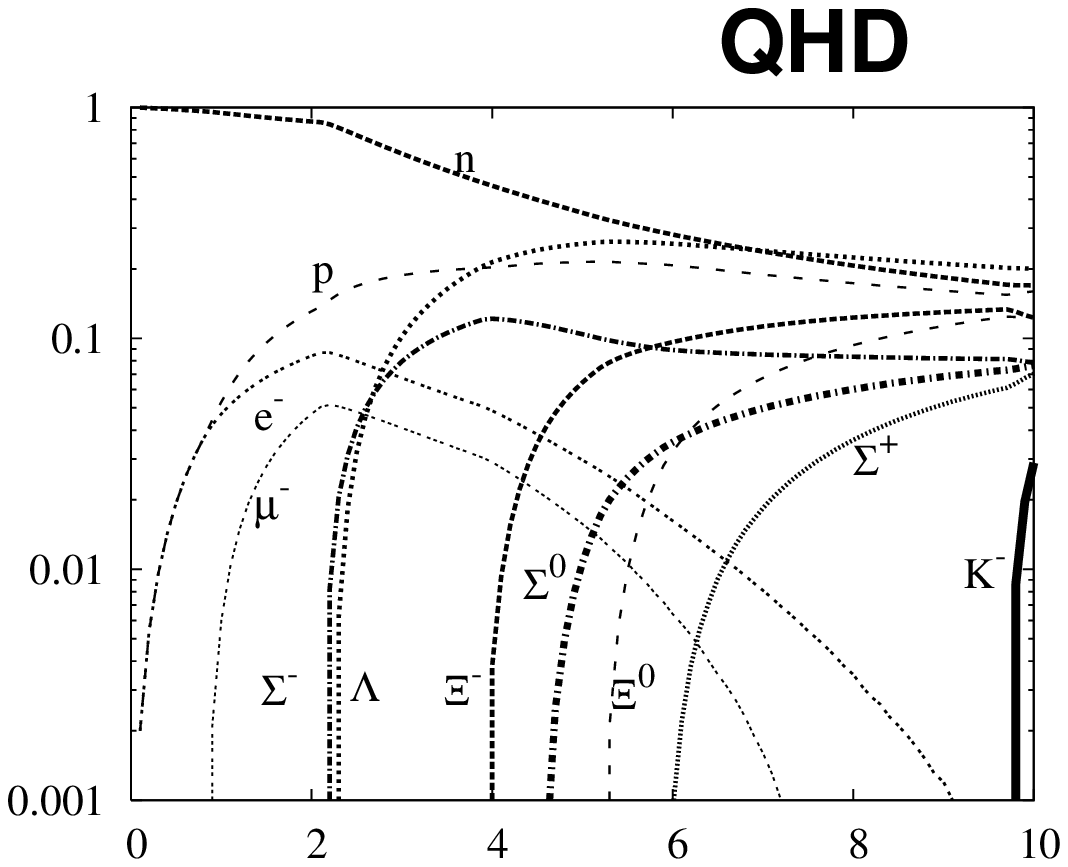, width=4.3cm}\\
\epsfig{file=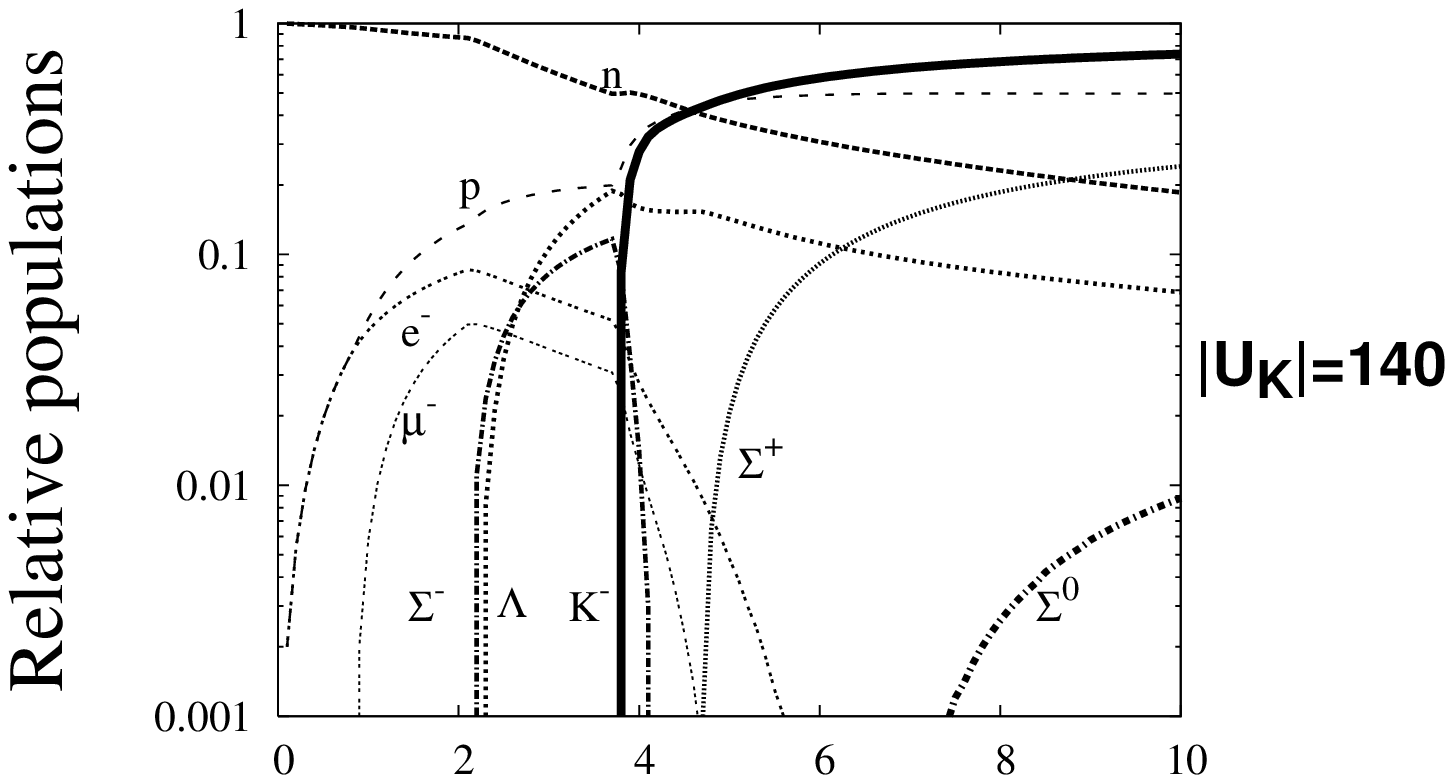, width=4.3cm}
\epsfig{file=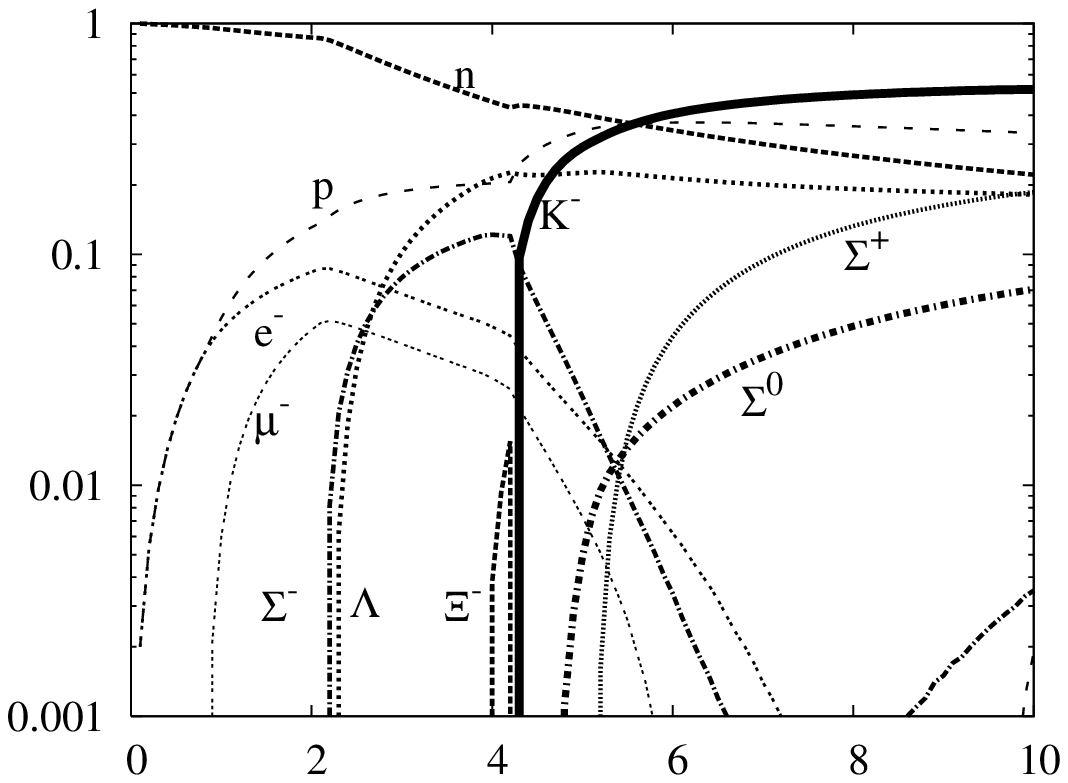, width=4.3cm}\\
\epsfig{file=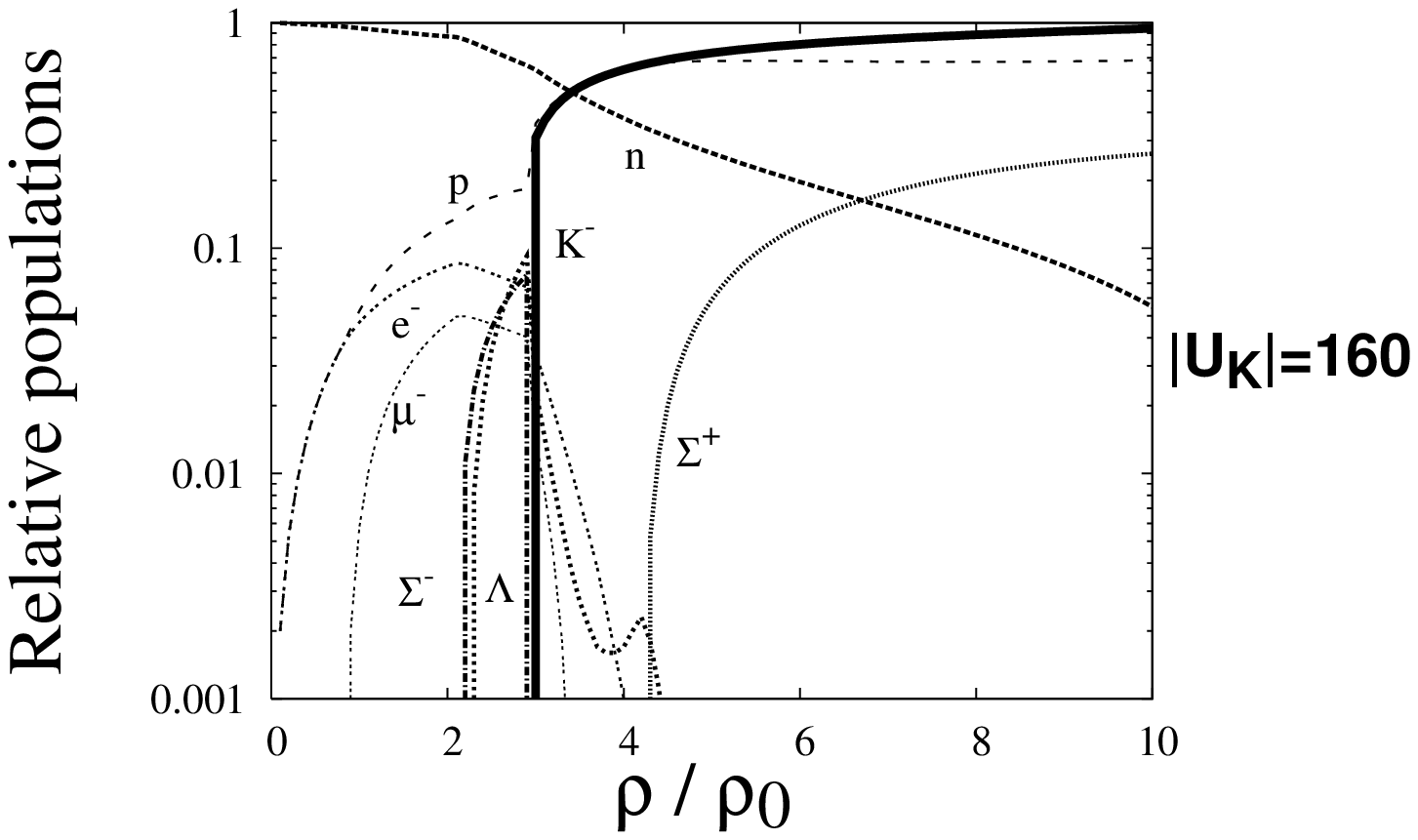, width=4.3cm}
\epsfig{file=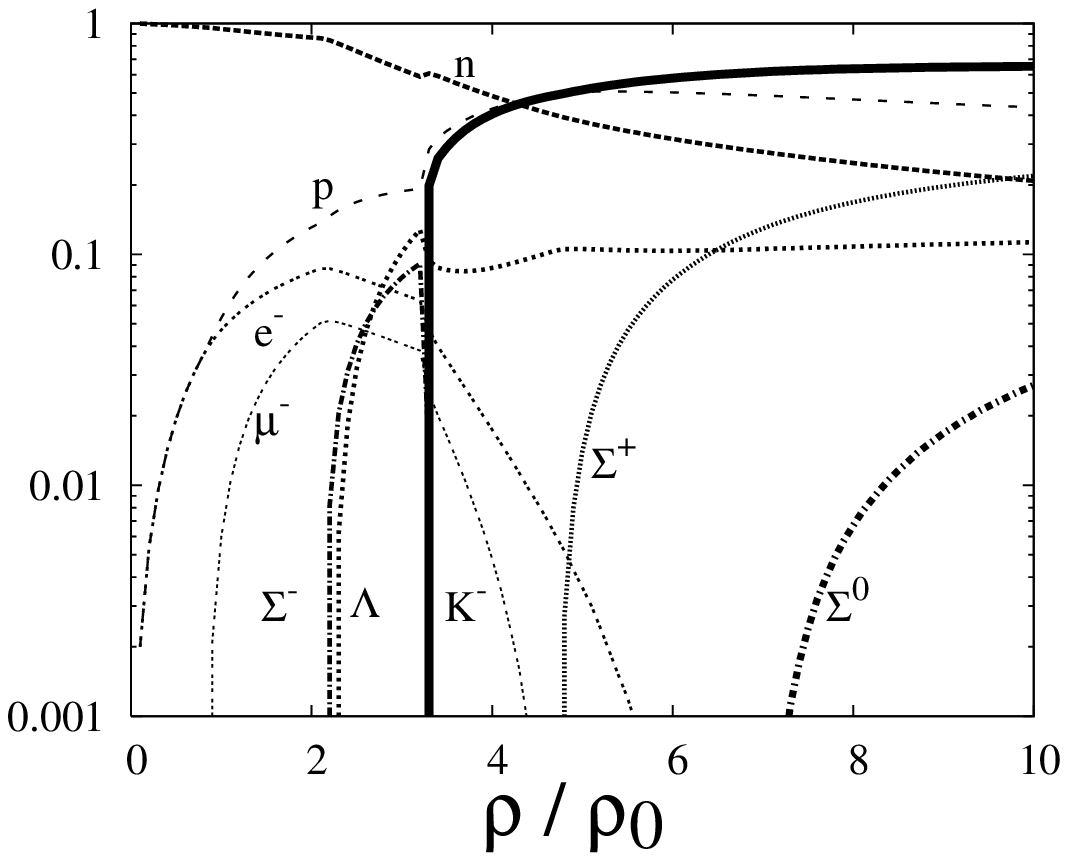, width=4.3cm}
\end{center}
\caption{The relative populations of particles in nuclear matter
calculated from the MQMC and QHD models are displayed 
in the left and right columns, respectively, for 
$U_{K^-} = -120$, $-140$ and $-160$ MeV.}
\label{fig:population}       
\end{figure}

A particle fraction is defined by the density of a particle divided by
the baryon density. Numerical results of the particle fraction is
shown in Fig.~\ref{fig:population}.
The onset density of the kaon condensation $\rho_{\rm crit}$ 
decreases as $|U_{K^-}|$ increases.
This is expected because a larger $g_{\sigma K}$ makes
$m^*_K$ smaller, which then causes the chemical equilibrium of kaons,
$\omega_K = \mu_n - \mu_p$ fulfilled at lower densities, where $\mu_{n(p)}$
is the chemical potential of the neutron (proton).
An interestring feature is that once the kaon is created,
the density of $K^-$ increases very rapidly and dominates 
the particle population over the hyperons and even the nucleons.
This can be partly attributed to the fact that the contribution from
the $\omega$-meson to the octet baryons 
is repulsive whereas the contribution from the $\omega$-meson
to the kaon is attractive.
The $\omega$-meson term in the energy of $K^-$ 
in Eq.~(\ref{eq:kaon-dispersion})  
has a negative sign and is thus attractive,
but it is repulsive for the chemical potential of the octet baryons.
The $\omega$-meson enhances the population of $K^-$
but suppresses baryons, and thus the kaon density increases rapidly.
In addition, due to the competition between
the negatively charged hyperons and $K^-$
in the charge neutrality condition, the negatively charged hyperons are highly
suppressed and in some cases are not even created at all
as soon as the kaon condensation sets in.
Positively charged hyperons, on the other hand, receive the
opposite effects from the kaon condensation, and
$\Sigma^+$ is created at lower densities as $|U_{K^-}|$ increases.
The proton density is also enhanced by large abundance of $K^-$,
which facilitates in turn the enhancement of $\Sigma^+$ population
through the chemical equilibrium condition
of the positively charged hyperons.

We find that the onset density of kaon condensation $\rho_{\rm crit}$ from
the MQMC model is lower than that from QHD.
For $U_{K^-} = -120, -140$, and $-160$ MeV, $\rho_{\rm crit}$
from MQMC are $5.9 \rho_0$, $3.8 \rho_0$ and $3.0 \rho_0$,
respectively, while they are
$9.8 \rho_0$, $4.3 \rho_0$ and $3.3 \rho_0$ in QHD.
The model dependence of $\rho_{\rm crit}$
becomes less pronounced for a larger $|U_{K^-}|$ value.
%

Kaon condensation leads to a phase transition 
from a pure hadronic phase to a kaon condensed one
with a mixed state of hadron and kaon phases.
To deal with such a mixed phase one needs to use Gibbs
conditions. However, for simplicity, we employ
Maxwell condition for the mixed phase, which is often used 
as an approximation to the Gibbs condition.
Though the Maxwell construction may not give us very accurate results
particularly for the EoS,
it can still provdie us with reasonable results for
the mass-radius relation of the neutron star.
\begin{figure}[b]
\begin{center}
\epsfig{file=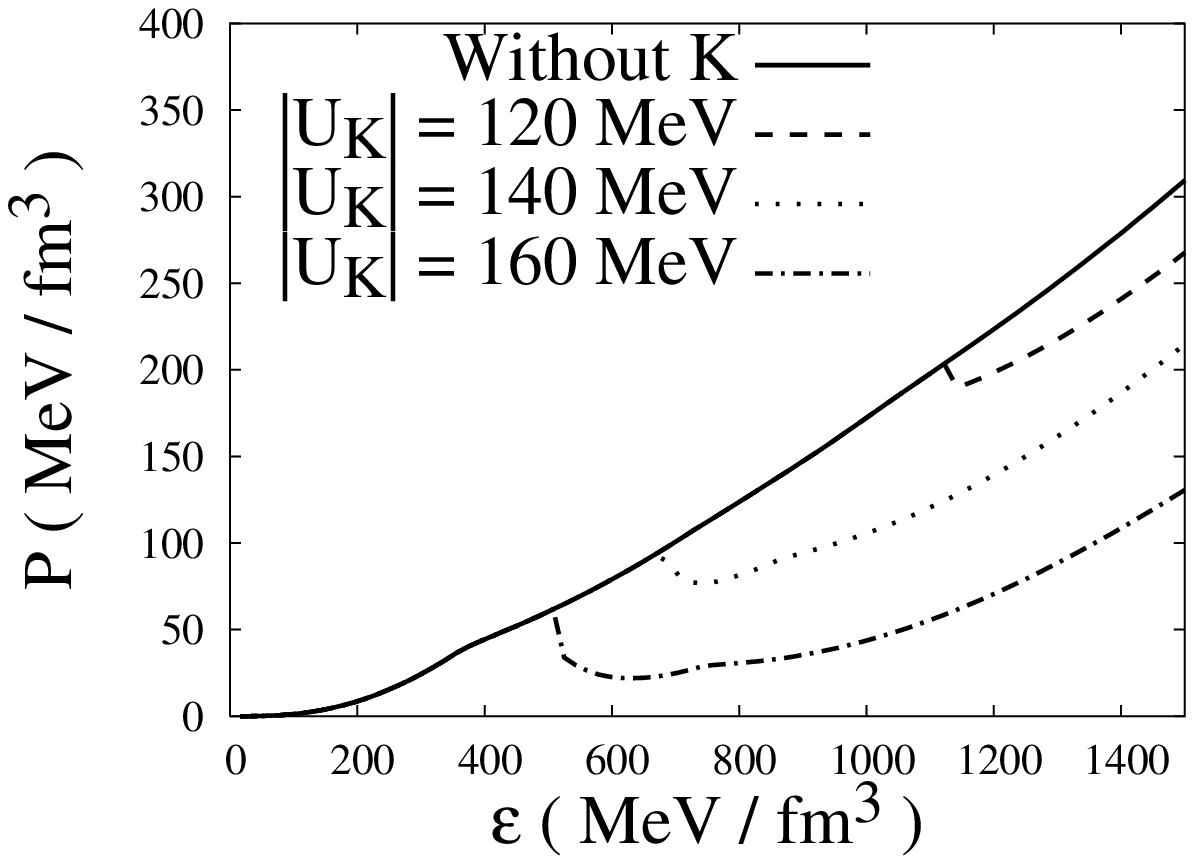, width=4.3cm}
\epsfig{file=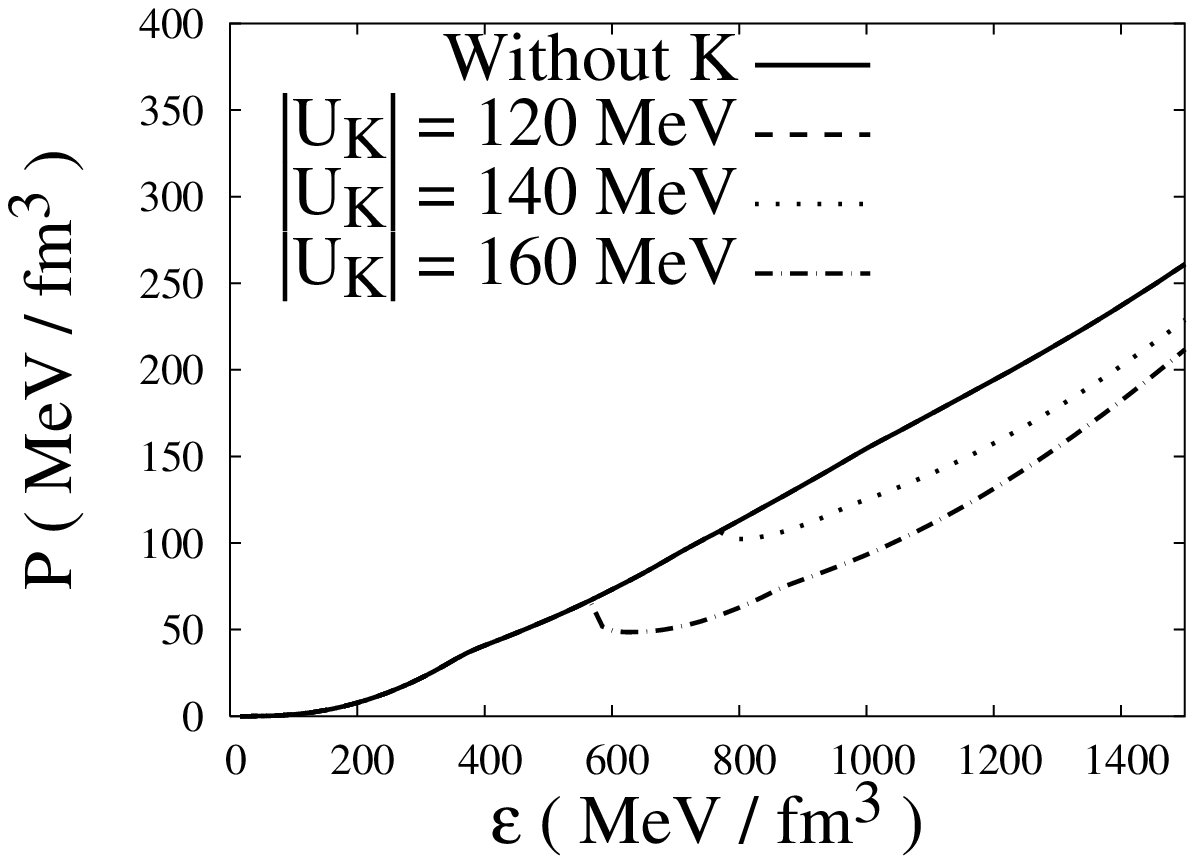, width=4.3cm}
\end{center}
\caption{The EoS without using any equilibrium condition in the mixed phase.
The left figure is from MQMC, and the right figure is from QHD.
The unstable region where the pressure decreases with increasing density
has to be remedied by either Maxwell or Gibbs condition.}
\label{fig:raw_eos}
\end{figure}

Fig.~\ref{fig:raw_eos} shows the EoS including kaon condensation, 
but calculated without using either Maxwell or Gibbs condition.
There is an instability region where pressure decreases as the density
increases. 
If such an instability exists in the interior of a neutron star, 
there is a high pressure region on top of a low pressure one.
The low pressure region cannot sustain the high pressure region, and
the unstable region will collapse, and the matter will be rearranged
to fix the instability.
Including such a rearrangement in the calculation of the EoS
is equivalent to implmenting Maxwell or Gibbs condition.
With Maxwell condition, the unstable part is replaced by a flat line,
which is determined by the equal area method. (The flat lines
are not plotted in Fig.~\ref{fig:raw_eos} to avoid too many lines
in one figure but should be obvious.)
From Fig.~\ref{fig:raw_eos} we can see that kaon condensation begins 
at a point where the EoS curve deviates from the solid curve
obtained without kaons. The kaon condensation  
makes the EoS softer than the one without it. 
As $|U_{K^-}|$ becomes larger, the EoS becomes softer.
By comparing the two solid curves in the left and in the right panels
we see that the EoS without kaons is softer with the QHD model,
but the effect of kaon condensation is more significant in the MQMC model.

\begin{figure}
\begin{center}
\epsfig{file=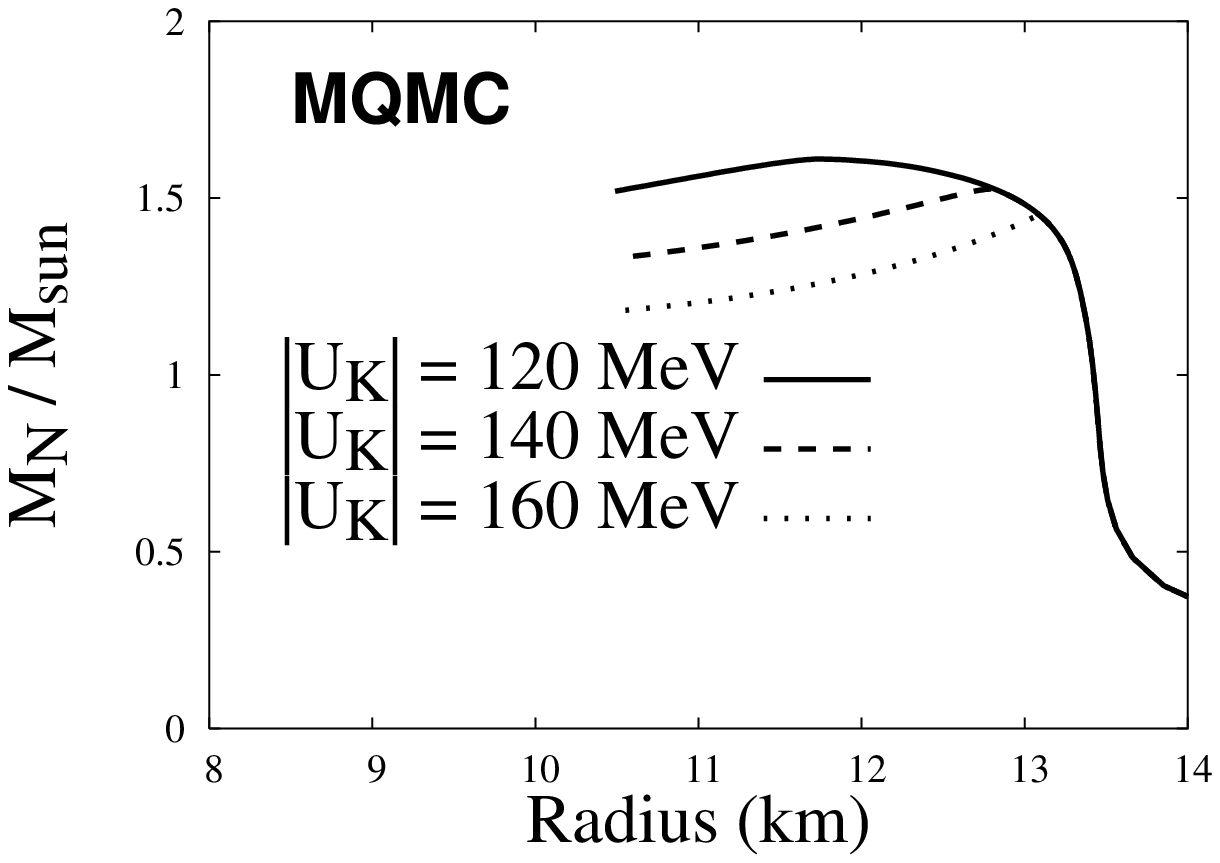,width=4.3cm}
\epsfig{file=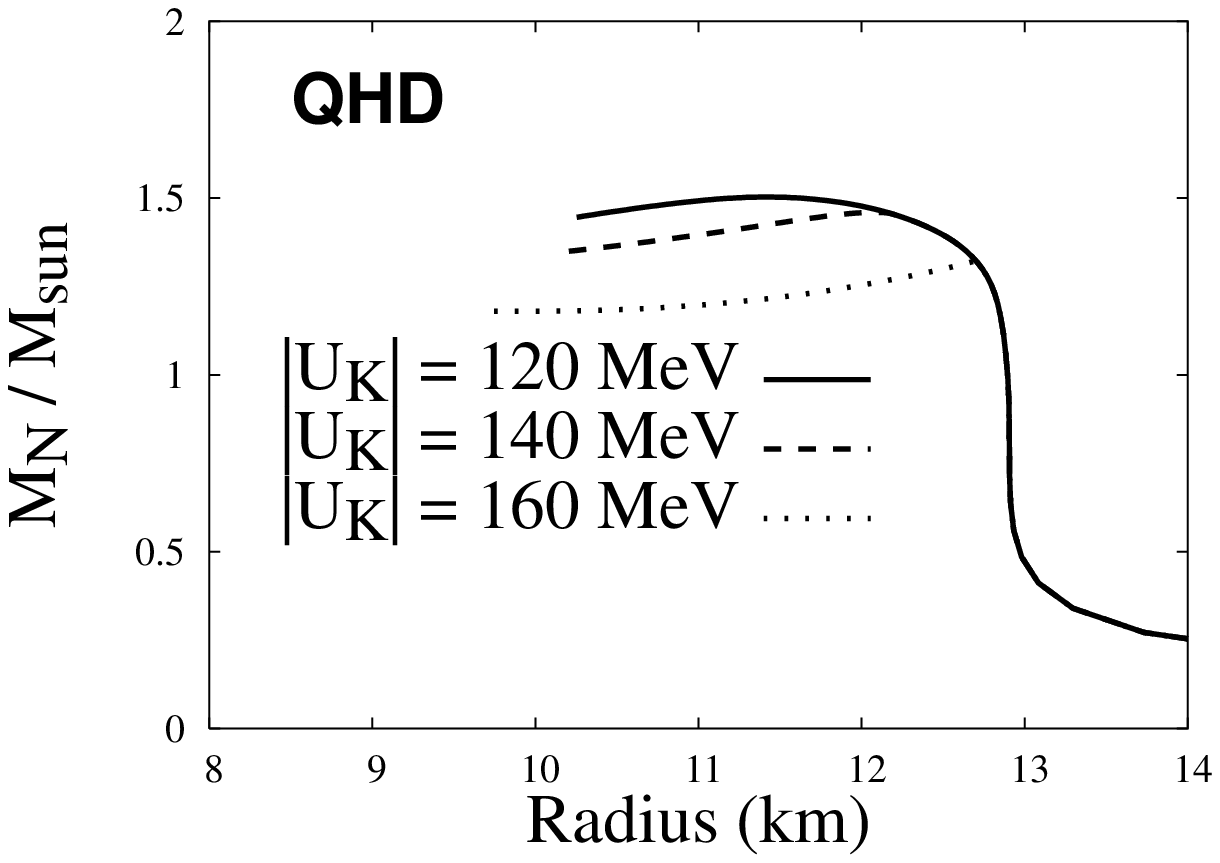,width=4.3cm}
\end{center}
\label{fig:mass-radius}
\caption{The mass-radius relation from the MQMC (left) and QHD (right) models.
Both figures are obtained with Maxwell construction.}
\end{figure}
Once the EoS is determined, it is straightforward to solve
the TOV equation to obtain the mass-radius relation of a neutron star.
Fig.~\ref{fig:mass-radius} shows the mass-radius curves calculated by
the MQMC (left) and QHD (right) models.
For $U_{K^-} = -120$ MeV, the effect of kaon condensation is 
rather small and negligible particularly for the QHD case.
The maximum mass of a neutron star with only nucleons is about $2 M_\odot$,
as mentioned in the Introduction.
When hyperons are created, the maximum mass of a neutron star 
decreases about $0.5 M_\odot$.
As the depth of optical potential for $K^-$ increases,
the effect of kaon condensation becomes more significant.
If $U_{K^-} = -160$ MeV, the maximum mass decreases by about 
$10 \sim 12$\% from the maximum mass without kaon condensation.
In the case of QHD the maximum mass with $U_{K^-} = -160$ MeV
becomes so small as $1.32 M_\odot$ that most of the observed masses in 
Ref.~\cite{tc-apj99} exceeds this (maximum) value.
This raises a question whether a very deep optical potential of
$K^-$ can be comptible with the existing observations of the neutron star mass.
We will discuss this again in the next section.

\begin{figure}
\begin{center}
\epsfig{file=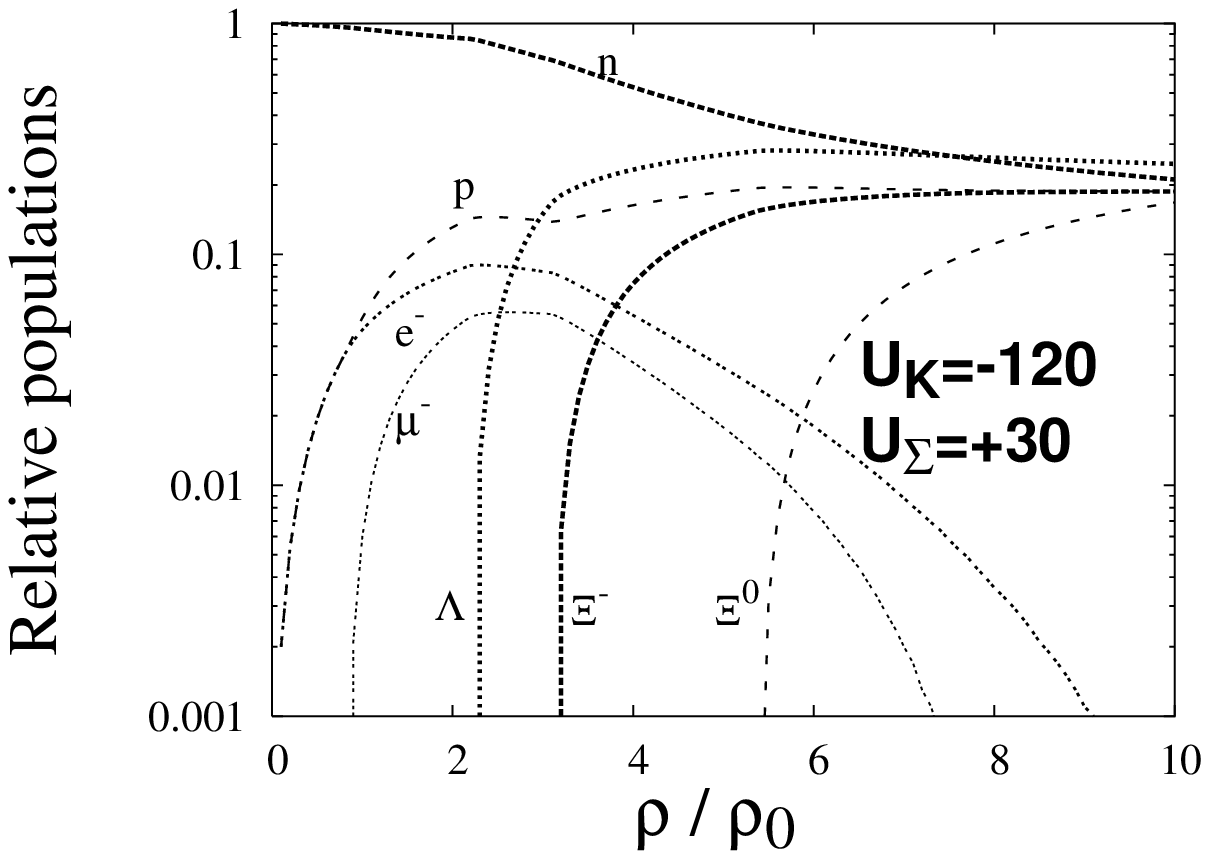,width=4.3cm}
\epsfig{file=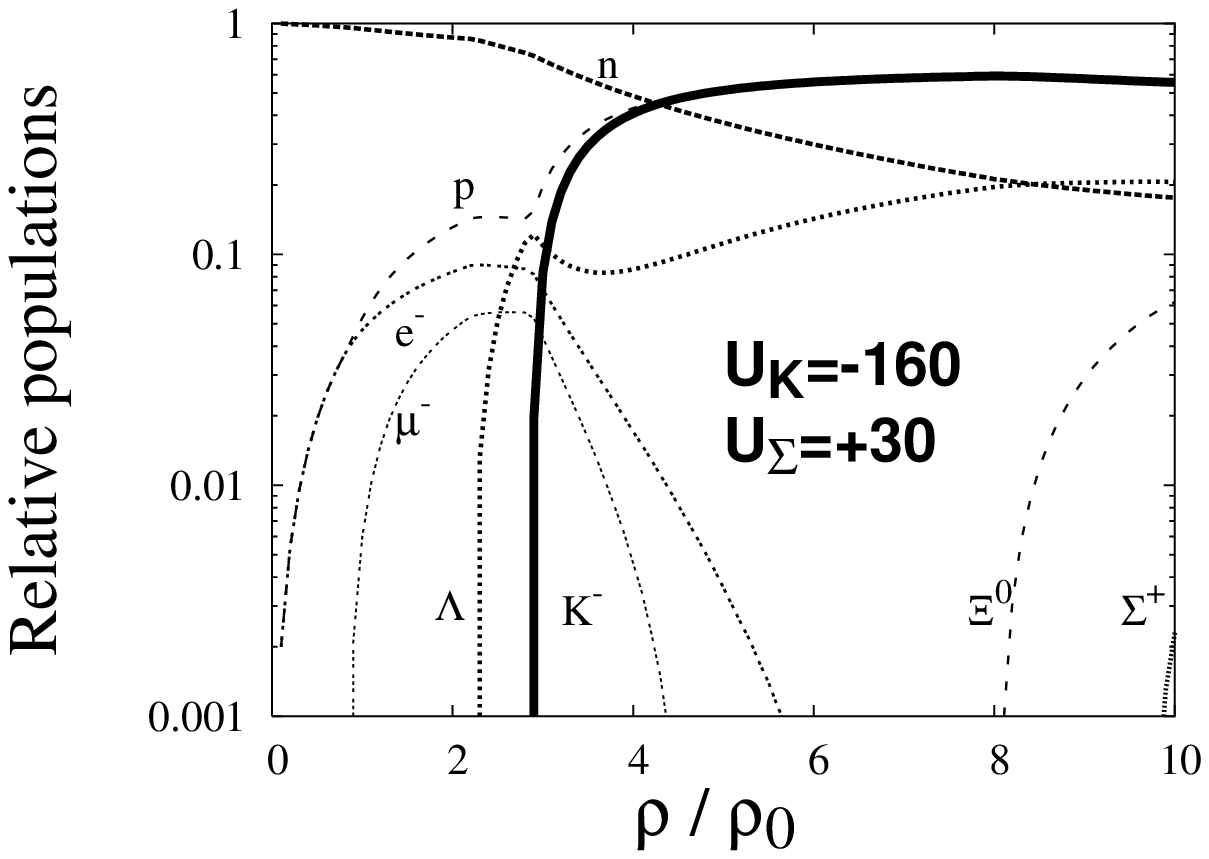,width=4.3cm}
\end{center}
\caption{The particle fraction with repulsive in-medium interaction
of $\Sigma$ hyperon in the QHD model. $U_\Sigma$ is fixed as $+30$ MeV,
and $U_{K^-}$ is chosen as $-120$ (left) and $-160$ (right) MeV.}
\label{fig:ppl_Sigma}
\end{figure}

One of the unsettled issues in hypernuclear physics is the
interaction of $\Sigma$ hyperon in nuclear medium. 
Use of plain quark counting rule for the hyperon-meson coupling constants 
gives us attractive optical potentials of $\Sigma$ hyperons 
in the range $-40 \sim -30$ MeV
at the saturation density. Some studies of $\Sigma$-hypernuclei, however,
indicates that the interaction of $\Sigma$ hyperon in medium is repulsive and
its optical potential at the saturation density takes a positive value
\cite{saha04}.
To take into account this possibility, we have repeated the calculations
using the QHD model with the repulsive $\Sigma$-hyperon 
optical potential of $+30$ MeV 
at the saturation density and show the results in Fig.~\ref{fig:ppl_Sigma}.
The particle compositions are displayed for $U_{K^-} = -120$ and 
$-160$ MeV. By comparing these two figures with the figures on the right panels
of Fig.~\ref{fig:population}, 
one can see that the repulsive $\Sigma$-hyperon interaction 
makes big changes in the population of hyperons, 
but the onset densities and the population of kaons are little affected.
The maximum masses of a neutron star with $U_\Sigma = +30$ MeV are
$1.52 M_\odot$, $1.48 M_\odot$ and $1.32 M_\odot$ for 
$U_{K^-} = -120$, $-140$ and $-160$ MeV, respectively.
These results deviate from those with attractive $\Sigma$-hyperon 
interaction only by about 1.4\% at most.
%
\section{Discussion}
\label{sec:conclusion}

Masses of some pulsars \cite{nice-apj05,page-06} 
that are more recently measured are reported
to lie much above the canonical value of Ref.~\cite{tc-apj99}.
Some of the large masses are in the range of $(1.7 - 2.1)\, M_\odot$, 
though the results depend on the binary systems. 
These values are even larger than the maximum mass of a neutron star
composed of only nucleons and hyperons (in short, a baryon star). 
If kaon condensate states are added to the baryonic state, 
it will soften the EoS and reduce the maximum mass further down.
Then one confronts the question: Are exotic states such as 
kaon condensation ruled out by large mass neutron stars?
To answer the question, we have to understand better
the interaction of particles and state of matter at densities above
the nuclear saturation. 
For instance, we don't know yet to what densities our mean field lagrangian 
models are valid.
If the large mass neutron star does exist, we have to have a mechanism
which makes the EoS at high densities stiffer than now.
Introducing hard core repulsion between nucleons can 
make the EoS stiff at high densities.
The effect of hard core has been considered in the high density/temperature
limit region \cite{ev1,ev2,ev3}, and it is shown that it makes the deconfined
quark phase the most stable state of matter at high density and/or temperature.
One can take into account such effects in the neutron star matter 
together with exotic states.
The investigation is in progress. \\

This work was supported by grant No.R01-2005-000-10050-0 
from the Basic Research
Program of the Korea Science \& Engineering Foundation.
Work of CHH was supported by Korea Research Foundation Grant funded
by Korea Government (MOEHRD, Basic Research Promotion Fund)
(KRF-2005-206-C00007).


\begin{thebibliography}{}
\bibitem{tc-apj99}
S. E. Thorsett and D. Chakrabarty,
Astrophys. J. {\bf 512}, 288 (1999).
\bibitem{kpe-prc95}
R. Knorren, M. Prakash and P. J. Ellis,
Phys. Rev. C {\bf 52}, 3470 (1995).
\bibitem{s-pal}
S. Pal, M. Hanauske, I. Zakout, H. St$\ddot {\rm o}$cker and W. Greiner,
Phys. Rev. C {\bf 60}, 015802 (1999).
\bibitem{nkg-prd92}
N. K. Glendenning,
Phys. Rev. D {\bf 46}, 1274 (1992).
\bibitem{Schaffner}
J. Schaffner-Bielich, V. Koch and M. Effenberger,
Nucl. Phys. {\bf A669}, 153 (2000).
\bibitem{oset20}
A. Ramos and E. Oset,
Nucl. Phys. {\bf A671}, 481 (2000).
\bibitem{Cieply}
A. Cieply, E. Friedman, A. Gal and J. Mares,
Nucl. Phys. {\bf A696}, 173 (2001).
\bibitem{gal94}
E. Friedman, A. Gal and C. J. Batty,
Nucl. Phys. {\bf A579}, 578 (1994).
\bibitem{Kaiser}
N. Kaiser, P.B. Siegel and W. Weise,
Nucl. Phys. {\bf A594}, 325 (1995).
\bibitem{Batty}
C. J. Batty, E. Friedman, A. Gal, Phys. Rep. {\bf 287}, 385 (1997).
\bibitem{mqmc}
X. Jin and B. K. Jennings,
Phys. Lett. {\bf B374}, 13 (1996);
Phys. Rev. C {\bf 54}, 1427 (1996).
\bibitem{sw-qhd}
B. D. Serot and J. D. Walecka,
Adv. Nucl. Phys. {\bf 16}, 1 (1986).
\bibitem{saha04}
P. K. Saha {\it et al.}, Phys. Rev. C {\bf 70}, 044613 (2004).
\bibitem{nice-apj05}
D. J. Nice {\it et al.}, Astrophys. J. {\bf 634}, 1242 (2005).
\bibitem{page-06}
D. Page and S. Reddy, Ann. Rev. Nucl. Part. Sci. {\bf 56}, 327 (2006).
\bibitem{ev1}
D. H. Rischke, M. I. Gorenstein, H. St\"{o}cker and W. Greiner,
Z. Phys. C {\bf 51}, 485 (1991).
\bibitem{ev2}
S. Kagiyama, A. Nakamura and T. Omodaka, 
Z. Phys. C {\bf 53}, 163 (1992).
\bibitem{ev3}
J. Cleymans, J. Stalnacke and E. Suhonen,
Z. Phys. C {\bf 55}, 317 (1992).
%
\end{thebibliography}
\end{document}